\documentclass[prl,aps,twocolumn,epsfig,superscriptaddress,showpacs]{revtex4}
\usepackage{bm}
\usepackage{amsfonts}
\usepackage[dvips]{graphicx}
\usepackage{mathrsfs}
\usepackage[intlimits]{amsmath}
\usepackage[colorlinks, citecolor=red]{hyperref}

\begin{document}

\title{Experimental Realization of Universal Geometric Quantum Gates with
Solid-State Spins}
\author{C. Zu$^{1}$, W.-B. Wang$^{1}$, L. He$^{1}$, W.-G. Zhang$^{1}$, C.-Y.
Dai$^{1}$, F. Wang$^{1}$, L.-M. Duan}
\affiliation{Center for Quantum Information, IIIS, Tsinghua University, Beijing 100084,
China}
\affiliation{Department of Physics, University of Michigan, Ann Arbor, Michigan 48109, USA}


\maketitle

\textbf{Experimental realization of a universal set of quantum logic gates
is the central requirement for implementation of a quantum computer. 
An all-geometric approach to quantum computation \cite%
{1,2} offered a paradigm for implementation where all the quantum gates are
achieved based on the Berry phases \cite{3} and their non-abelian
extensions, the holonomies \cite{5}, from geometric transformation of
quantum states in the Hilbert space \cite{5a}. Apart from its fundamental
interest and rich mathematical structure, the geometric approach has some
built-in noise-resilient features \cite{1,2,6,7}. On the experimental side,
geometric phases and holonomies have been observed using
nuclear magnetic resonance with thermal ensembles of liquid molecules \cite%
{8,10}, however, such systems are known to be non-scalable for quantum
computing \cite{11}. There are proposals to implement geometric quantum
computation in scalable experimental platforms such as trapped ions \cite{12}%
, superconducting qubits \cite{13}, or quantum dots \cite{14}, and a recent
experiment has realized geometric single-bit gates with the superconducting
system \cite{15}. Here, we report the first experimental realization of a
universal set of geometric quantum gates with solid-state spins of the diamond
defects. The diamond defects provide a scalable experimental platform \cite{26,16b,16} with
the potential for room-temperature quantum computing \cite{16b,16,17,18b},
which has attracted strong interest in recent years \cite{18}. Based on
advance of coherent control in this system \cite{26,16b,16,17,18b,18}, our experiment shows that
all-geometric and potentially robust quantum computation can be realized
with solid-state spin qubits.}

Under adiabatic cyclic evolution, a non-degenerate eigenstate of a quantum
system acquires a phase factor, which has a dynamical component proportional
to the time integral of the eigenenergy and a geometric component determined
by the global property of the evolution path. This geometric phase, first
discovered by Berry \textbf{\cite{3}}, has found connection with many
important physics phenomena \textbf{\cite{19}}. If the system has degenerate
eigenstates, the Berry's phase is replaced by an geometric unitary operator
acting on the degenerate subspace, termed as holonomy from the differential
geometry. The holonomies are in general non-commutable with each other. In
the proposal of geometric quantum computation \textbf{\cite{1,2}}, such
holonomies are exploited to realize a universal set of quantum gates,
compositions of which then can fulfill arbitrary quantum computation tasks.
As holonomies are determined by global geometric properties, geometric
computation is more robust to certain control errors \textbf{\cite{1,2,6,7}}%
. Implementation of geometric quantum computation has been proposed in
several qubit systems \textbf{\cite{12,13,14}}, however, it remains
experimentally challenging to realize a universal set of gates all by
holonomies, because of the requirements of slow adiabatic evolution and a
complicated level structure.

In the recent proposal of non-adiabatic geometric quantum computation
\textbf{\cite{6,20}}, universal quantum gates are constructed fully by
geometric means without requirement of the adiabatic condition, thereby
combining speed with universality. Under a cyclic evolution of the system
Hamiltonian $H\left( t\right) $ (with $H\left( \tau \right) =H\left(
0\right) $), let $\left\vert \xi _{l}\left( t\right) \right\rangle $ ($%
l=1,2,\cdots ,M$) denote instantaneous orthonormal bases (moving frames)
which coincide with the basisvectors $\left\vert \xi _{l}\right\rangle $ of
the computational space $C$ at $t=0,\tau $ with $\left\vert \xi _{l}\left(
\tau \right) \right\rangle =\left\vert \xi _{l}\left( 0\right) \right\rangle
=\left\vert \xi _{l}\right\rangle $. The evolution operator $U\left( \tau
\right) $ on the basis states $\left\vert \xi _{l}\right\rangle $ has two
contributions: a dynamic part and a fully geometric part \textbf{\cite{6}}.
If the parallel-transport condition $\left\langle \xi _{l}\left( t\right)
\right\vert H\left( t\right) \left\vert \xi _{l^{\prime }}\left( t\right)
\right\rangle =0$ is satisfied for any $l,l^{\prime }$ at any time $t$, the
dynamic contribution becomes identically zero, and $U\left( \tau \right) $
is given by
\begin{equation}
U\left( \tau \right) =T\exp \left[ i\int_{0}^{\tau }Adt\right] ,
\end{equation}%
where $T$ denotes the time-ordered integration and $A=\left[ A_{ll^{\prime }}%
\right] =\left[ \left\langle \xi _{l}\left( t\right) \right\vert i\partial
_{t}\left\vert \xi _{l^{\prime }}\left( t\right) \right\rangle \right] $
represents the $M\times M$ connection matrix \textbf{\cite{6}}. The form of $%
U\left( \tau \right) $ is identical to the Wilczek-Zee holonomy in the
adiabatic case \textbf{\cite{5,6}}.

\begin{figure}[tbp]
\includegraphics[width=8.5cm,height=10cm]{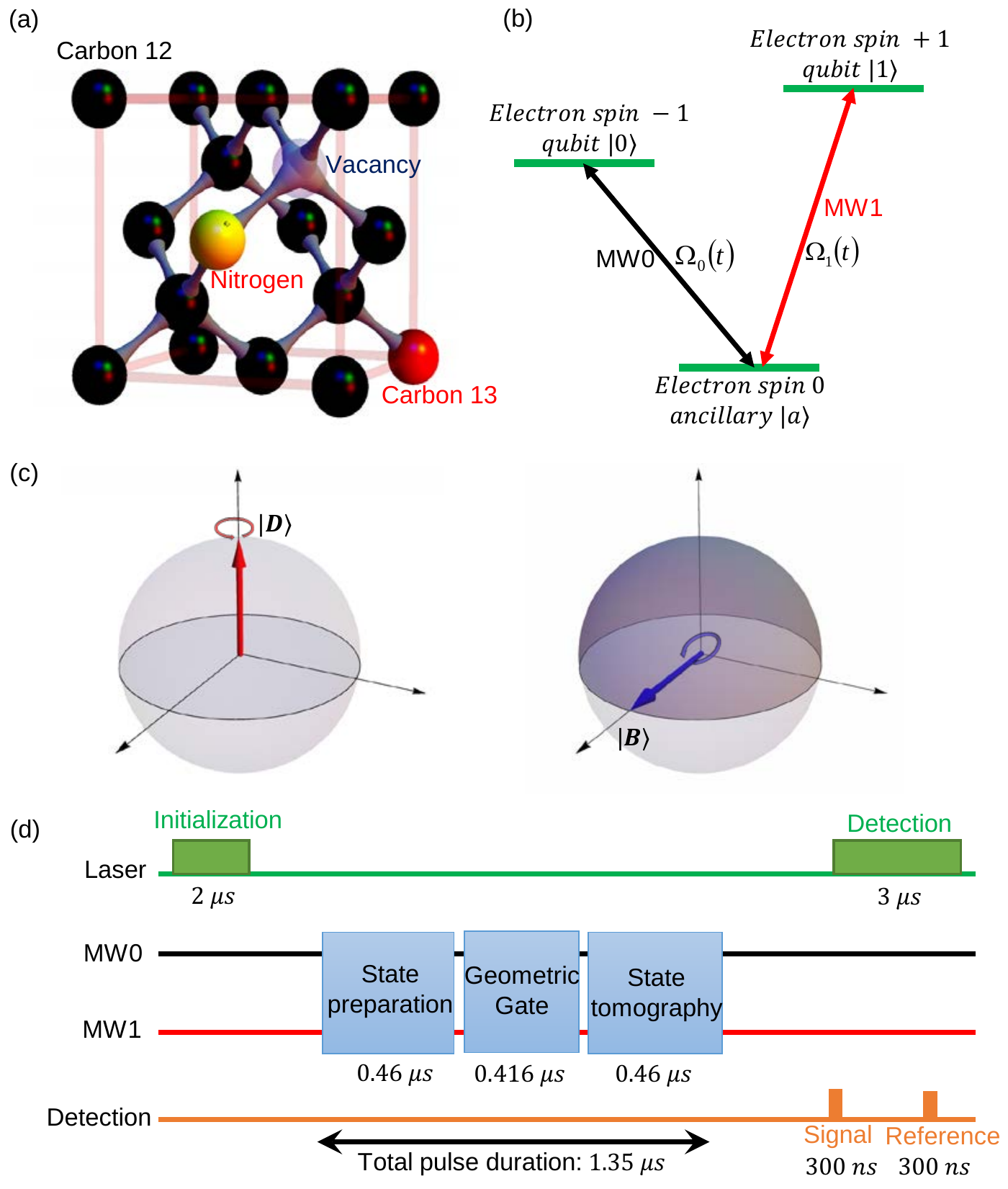}
\caption[Fig. 1 ]{\textbf{Geometric gates in a diamond nitrogen vacancy center.}
{\bf a}, Illustration of a nitrogen vacancy (NV) center in a
diamond with a proximal $C^{13}$ atom. {\bf b}, Encoding of a qubit in the
spin-triplet ground state of the NV center and the microwave coupling
configuration. The electron spin $0$ state provides an ancillary level $%
\left\vert a\right\rangle$ for geometric manipulation of the qubit. {\bf c},
A geometric picture of the holonomic gates. Under a
cyclic Hamiltonian evolution, the dark $\left\vert D\right\rangle$ (bright $%
\left\vert B\right\rangle$) state rotates by $2\protect\pi$ along the north
pole (equator) of the Bloch sphere, acquiring a geometric phase of $0$ ($%
\protect\pi$) given as half of the swept solid angle. When we choose
different forms of the dark and bright states by controlling parameters in
the Hamiltonian, this state-dependent geometric phase leads to the
corresponding holonomic gates. {\bf d}, The time sequence for
implementation and verification of single-qubit geometric gates. }
\end{figure}

Our experiment realizes a universal set of quantum gates all by use of the
nonadiabatic holonomies \textbf{\cite{6}}. Single-bit gates, together with
entangling controlled-NOT (CNOT) operation, are universal for quantum
computation. Our realization is based on control of electron and nuclear spins in a diamond
nitrogen-vacancy (NV) center that form effectively a quantum register \textbf{\cite{18}}. To
realize the single-bit geometric gates, we manipulate the electron spin
states of a NV center (Fig. 1a) in a synthetic diamond at room temperature
(see Methods for description of the experimental setup). The NV\ center has
a spin-triplet ground state. We take the Zeeman components $%
\left\vert m=-1\right\rangle \equiv \left\vert 0\right\rangle $ and $%
\left\vert m=+1\right\rangle \equiv \left\vert 1\right\rangle $ as the qubit
basis states and use $\left\vert m=0\right\rangle \equiv \left\vert
a\right\rangle $ as an ancillary level for geometric manipulation of the
qubit. The spin state is initialized through optical pumping to the $%
\left\vert m=0\right\rangle $ level and read out by distinguishing different
fluorescence levels of the states under illumination of a short green laser
pulse \textbf{\cite{18}} (see Methods for calibration of fluorescence levels
of different states). We apply a magnetic field of $451$ G along the NV axis
using a permanent magnet. Under this field, the nearby nuclear spins are
polarized by optical pumping \textbf{\cite{22}}, enhancing the coherence
time of the electron spin.

The transitions from the qubit states $\left\vert 0\right\rangle ,\left\vert
1\right\rangle $ to the ancillary level $\left\vert a\right\rangle $ are
coupled by microwave pulses controlled through an arbitrary waveform
generator (AWG), with Rabi frequencies $\Omega _{0}\left( t\right) $, $%
\Omega _{1}\left( t\right) $, respectively (Fig. 1b). We vary the amplitude $%
\Omega \left( t\right) =\sqrt{\Omega _{0}^{2}+\Omega _{1}^{2}}$ but fix the
ratio $\Omega _{1}/\Omega _{0}=e^{i\varphi }\tan \theta $ to be constant.
The Hamiltonian for the coupling between these three levels takes the form
\begin{equation}
H_{1}\left( t\right) =\hbar \Omega \left( t\right) \left[ \left( \cos \theta
\left\vert 0\right\rangle +e^{i\varphi }\sin \theta \left\vert
1\right\rangle \right) \left\langle a\right\vert +H.c.\right]
\end{equation}%
where $\hbar $ is the Planck constant divided by $2\pi $ and $H.c.$ denotes
the Hermitian conjugate. Define the bright state as $\left\vert
B\right\rangle =\cos \theta \left\vert 0\right\rangle +e^{i\varphi }\sin
\theta \left\vert 1\right\rangle $ and the dark state as $\left\vert
D\right\rangle =-e^{-i\varphi }\sin \theta \left\vert 0\right\rangle +\cos
\theta \left\vert 1\right\rangle $. When $\Omega \left( t\right) $ makes a
cyclic evolution with $\Omega \left( 0\right) =\Omega \left( \tau \right) =0$%
, the bright state evolves as $\left\vert B\left( t\right) \right\rangle
=e^{i\alpha \left( t\right) }\left[ \cos \alpha \left( t\right) \left\vert
B\right\rangle +\sin \alpha \left( t\right) \left\vert a\right\rangle \right]
$, where $\alpha \left( t\right) \equiv \int_{0}^{t}\Omega \left( t^{\prime
}\right) dt^{\prime }$, while the dark state remains unchanged. After a
cyclic evolution with $\alpha \left( \tau \right) =\pi $, the bright (dark)
state picks up a geometric phase of $\pi $ ($0$), respectively, as
illustrated in Fig. 1c. We take the moving frame as $\left\vert \xi
_{0}\left( t\right) \right\rangle =\cos \theta \left\vert B\left( t\right)
\right\rangle -e^{i\varphi }\sin \theta \left\vert D\right\rangle $, $%
\left\vert \xi _{1}\left( t\right) \right\rangle =e^{-i\varphi }\sin \theta
\left\vert B\left( t\right) \right\rangle +\cos \theta \left\vert
D\right\rangle $, which makes a cyclic evolution with $\left\vert \xi
_{l}\left( 0\right) \right\rangle =\left\vert \xi _{l}\left( \tau \right)
\right\rangle =\left\vert l\right\rangle $ $\left( l=0,1\right) $.\ For this
evolution, one can easily check that the condition $\left\langle \xi
_{l}\left( t\right) \right\vert H\left( t\right) \left\vert \xi _{l^{\prime
}}\left( t\right) \right\rangle =0$ is always satisfied, so there is no
dynamic contribution to the evolution operator $U\left( \tau \right) $
\textbf{\cite{6}}. Using the expression (1), we find the holonomy%
\begin{equation}
U\left( \tau \right) =\left[
\begin{array}{cc}
-\cos \left( 2\theta \right) & -e^{i\varphi }\sin \left( 2\theta \right) \\
-e^{-i\varphi }\sin \left( 2\theta \right) & \cos \left( 2\theta \right)%
\end{array}%
\right]
\end{equation}%
under the computational basis $\left\{ \left\vert 0\right\rangle ,\left\vert
1\right\rangle \right\} $.

\begin{figure}[tbp]
\includegraphics[width=8.5cm,height=8cm]{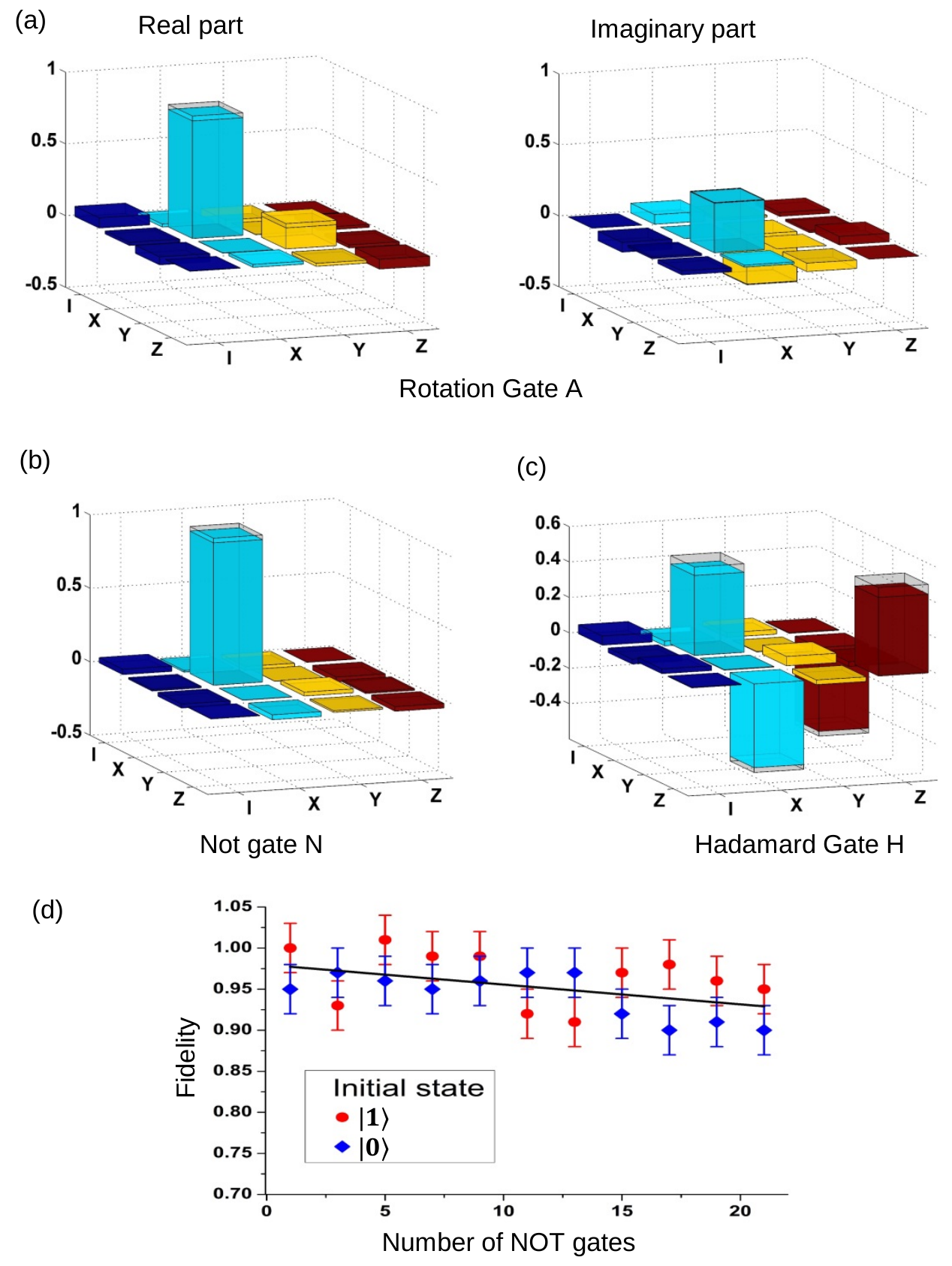}
\caption[Fig. 2 ]{\textbf{Experimental results for single-bit geometric gates.}
The measured process matrix elements for the rotation gate
$A$ ({\bf a}), the NOT gate $N$ ({\bf b}), and the Hadamard gate $H$ ({\bf c}). The measured
tiny imaginary parts of the process matrices for the NOT and the Hadamard
gates are not shown. The hollow caps in these figures denote the
corresponding matrix elements for the ideal gates. {\bf d}, The measured
fidelities of the final states compared with the ideal output (error bars denote s.d.) after
application of a sequence of the geometric NOT gates, with the initial state
taken as $\left\vert 0 \right\rangle $ and $\left\vert 1 \right\rangle $,
respectively. By fitting the data under the assumption of independent error
for each gate, we obtain the error induced by each NOT gate at $(0.24
\pm 0.06)\% $. }
\end{figure}

We evolve the Rabi frequencies $\Omega _{i}\left( t\right) $ along three
different loops, with the parameters $\left( \theta ,\varphi \right) $
chosen respectively as $\left( 3\pi /4,0\right) $, $\left( 3\pi /4,\pi
/8\right) $, $\left( 5\pi /8,0\right) $. The three geometric gates resulting
from these cyclic evolutions are denoted by the NOT gate $N$, the rotation
gate $A$, and the Hadamard gate $H$, respectively. The combination of the
gates $N$ and $A$ gives the well-known $\pi /8$-gate $T=NA$, which, together
with the Hadamard gate $H$, make a universal set of single-bit gates. To
characterize these geometric gates, we use quantum process tomography by
preparing and measuring the qubit in different bases \cite{23b}, with the
time sequence shown in Fig. 1d. The matrix elements for each process are
shown in Fig. 2a-2c, which are compared with the corresponding elements of
the ideal gates. From the process tomography (see Methods), we find the
process fidelity $F_{P}=\left( 96.5\pm 1.9\right) \%$, $\left( 96.9\pm
1.5\right) \%$, $\left( 92.1\pm 1.8\right) \%$ respectively for the $N$, $A$%
, and $H$ gates. The major contribution to the infidelity actually comes
from the state preparation and detection error in quantum process
tomography. To measure the intrinsic gate error, we concatenate a series of
gates and examine the fidelity decay as the number of gates increases
\textbf{\cite{18b}}. As an example, we show in Fig. 2d the fidelity decay by
concatenating the NOT gates. From the data, we find the intrinsic error per
gate is about $0.24\%$. This can be compared with the $1\%$ error rate for
the dynamic NOT\ gate using optimized pulses\textbf{\ }by the same method of
measurement \textbf{\cite{18b}}. The achieved high fidelity indicates that
the geometric manipulation is indeed resilient to control errors.

\begin{figure}[tbp]
\includegraphics[width=8.5cm,height=8cm]{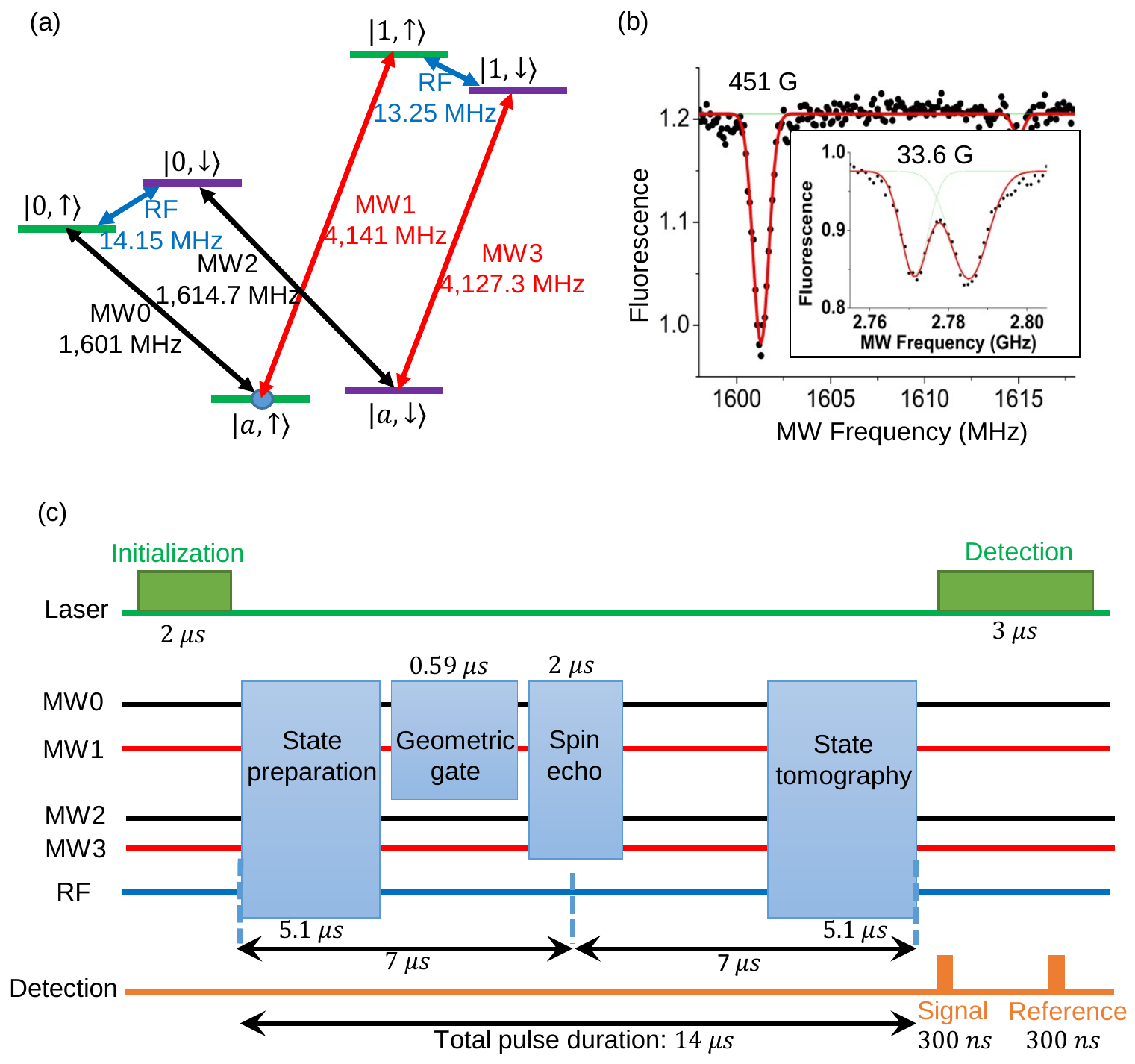}
\caption[Fig. 3 ]{ \textbf{Level scheme and pulse sequence for the geometric CNOT gate.}
{\bf a}, The level structure of the electron and the nuclear
spins for the geometric CNOT gate and the microwaves and RF coupling
configuration. {\bf b}, Optically detected magnetic resonance (ODMR) spectroscopy
by measuring the fluorescence level while scanning the frequency of the
microwave that couples to the electron spin $0$ to $1$ transition. The two
dips at $33.6$ G magnetic field (shown in the insert) represent the
hyperfine splitting caused by the unpolarized nuclear spin. The very
asymmetric dips at $451$ G field indicates that the nuclear spin has been
polarized. {\bf c}, The time sequence for implementation and verification of the
geometric CNOT gate between the electron and the nuclear spins. The CNOT
gate is implemented by applying MW0 and MW1 pulses simultaneously. The other
microwaves (MW0, MW1, MW2, and MW3) are used for implementation of a spin
echo to increase the spin coherence time. To verify the CNOT gate, we use a
combination of the microwave and the RF pulses to prepare various initial
superposition states and measure the final output in different bases
through quantum state tomography. }
\end{figure}

To realize the geometric quantum CNOT\ gate, we use one nearby $C^{13}$
nuclear spin as the control qubit (with the basisvectors $\left\vert
\uparrow \right\rangle ,\left\langle \downarrow \right\vert $) and the NV
center electron spin as the target qubit \cite{22a}. Both the electron spin
and the nuclear spin are polarized through optical pumping under the $451$ G
magnetic field, which is confirmed by the optically detected magnetic
resonance (ODMR) spectroscopy shown in Fig. 3b. The spins are interacting
with each other through hyperfine and dipole couplings, and the resultant
level configuration is shown in Fig. 3a. By applying state-selective
microwave (MW) and radio-frequency (RF) pulses, we can couple different
levels. In particular, with the MW0 and MW1 pulses with Rabi frequencies $%
\Omega _{0}\left( t\right) $, $\Omega _{1}\left( t\right) $, we have the
following coupling Hamiltonian
\begin{equation}
H_{2}\left( t\right) =\hbar \Omega \left( t\right) \left[ \left( \left\vert
0,\uparrow \right\rangle -\left\vert 1,\uparrow \right\rangle \right)
\left\langle a,\uparrow \right\vert +H.c.\right] /\sqrt{2},
\end{equation}%
where we have fixed the ratio $\Omega _{1}/\Omega _{0}=-1$. Under a cyclic
evolution of $\Omega \left( t\right) $ with $\int_{0}^{\tau }\Omega \left(
t\right) dt=\pi $, we find the holonomy $U\left( \tau \right) =\left\vert
\uparrow \right\rangle \left\langle \uparrow \right\vert \otimes
N+\left\vert \downarrow \right\rangle \left\langle \downarrow \right\vert
\otimes I$ using the formula (1), where $I$ denotes the $2\times 2$ unit
matrix. This achieves exactly the quantum CNOT\ gate.

\begin{figure}[tbp]
\includegraphics[width=8.5cm,height=6cm]{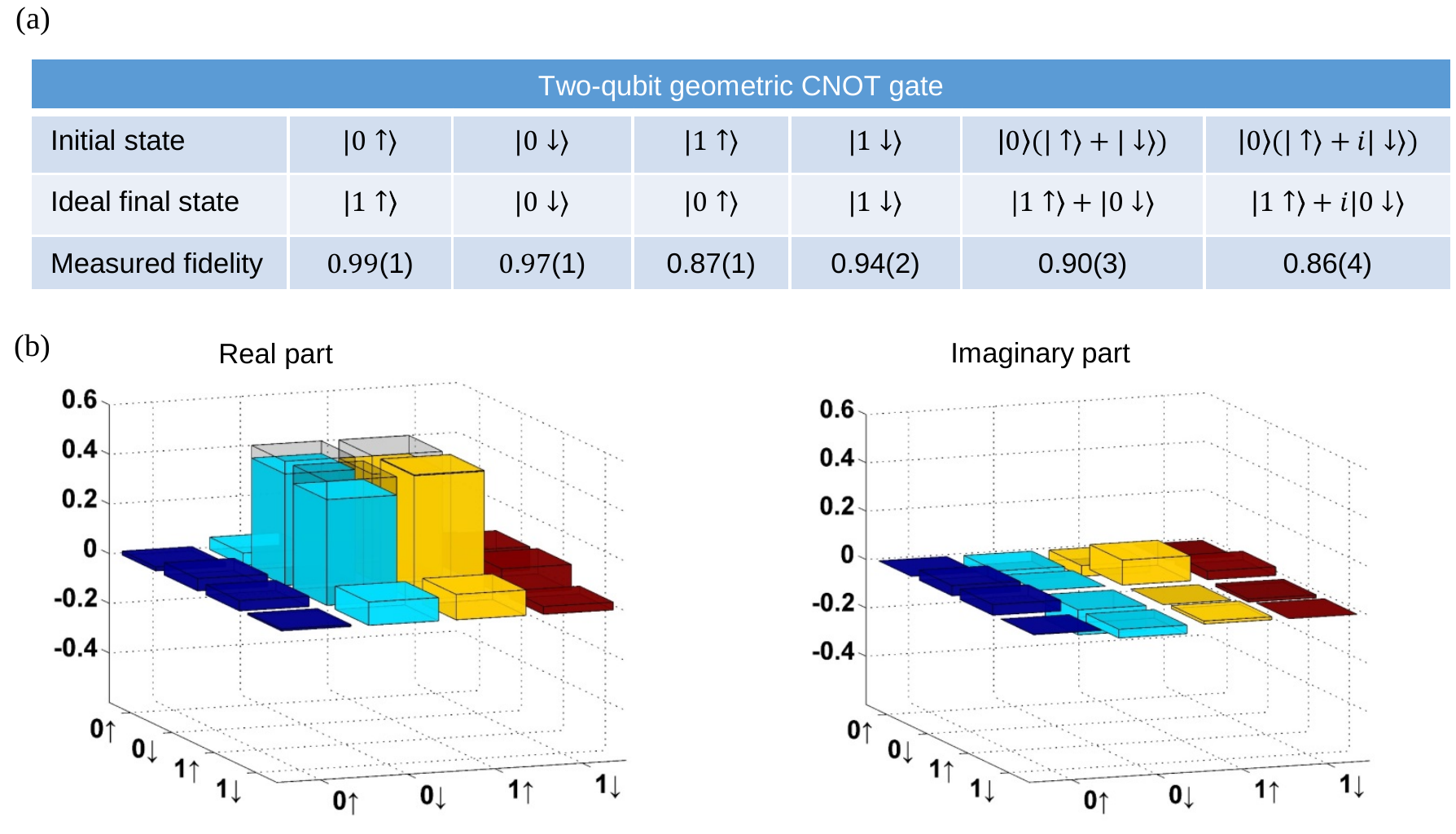}
\caption[Fig. 4 ]{\textbf{Experimental results for the geometric CNOT gate.}
{\bf a}, Measured output state fidelities of the geometric
CNOT gate under a few typical input states, where the number in the bracket
represents the error bar (s.d.) in the last digit. {\bf b}, The matrix elements of the
output density operator reconstructed through quantum state tomography when
the geometric CNOT is applied to the product state $\left\vert
0\right\rangle \left( \left\vert \uparrow \right\rangle +\left\vert
\downarrow \right\rangle \right) /\protect\sqrt{2}$. The hollow caps denote
the matrix elements for the ideal output state under a perfect gate.}
\end{figure}

To characterize the geometric CNOT\ gate, we apply the gate to the qubit
basis states as well as their superpositions, and measure the fidelity of
the final states compared with the ideal outputs through the quantum state
tomography \cite{23b}. The superposition of the nuclear spin states required
for state preparation and measurement is generated through RF pulses,
which take longer times compared with the microwave pulses due to the much
smaller magnetic moment of the nuclear spin. The electron spin decoherence
is significant during the slow RF pulses. To correct that, we apply a Hahn
spin echo in the middle of the whole operation with the time sequence shown
in Fig. 3c. The measured state fidelities are listed in Fig. 4a under
typical input states. As a hallmark of the entangling operation, the
geometric CNOT\ gate generates entanglement from the initial product state.
As an example, for the input state $\left\vert 0\right\rangle \otimes \left(
\left\vert \uparrow \right\rangle +\left\vert \downarrow \right\rangle
\right) \ $(unnormalized), the matrix elements of the output density
operator are shown in Fig. 4b, with a measured entanglement fidelity of $%
\left( 90.2\pm 2.5\right) \%$ and concurrence of $0.85\pm 0.05$, which
unambiguously confirms entanglement \cite{11}.

Our experimental realization of a universal set of holonomic gates with
individual spins paves the way for all-geometric quantum computation
in a solid-state system. The electron and nuclear spins of different
NV\ centers can be wired up quantum mechanically to form a scalable network
of qubits through, e.g., the direct dipole interaction \cite{16b,17}, the
spin-chain assisted coupling by the nitrogen dopants \cite{16,23}, or the
photon-mediated coupling \cite{26,24,25}. The technique employed here for
geometric realization of universal gates may also find applications in other
scalable experimental systems, such as trapped ions or superconducting
qubits. The geometric phase is closely related to the topological phase \cite%
{27b,27}, and the demonstration of gates all by holonomies is an important
step towards realization of topological computation \cite{27}, the most
robust way of quantum computing.

\textbf{Acknowledgements} We thank M. Lukin's group for helpful discussions.
This work was supported by the National Basic Research Program of China
2011CBA00300 (2011CBA00302) and the quantum information project from the Ministry of Education
of China. LMD acknowledges in addition support from the IARPA MUSIQC program, the AFOSR and the ARO MURI
program.

\textbf{Author Contributions} L.M.D. conceived the experiment and supervised
the project. C.Z., W.B.W., L.H., W.G.Z., C.Y.D., F.W. carried out the
experiment. L.M.D. and C.Z. wrote the manuscript.

\textbf{Author Information} Reprints and permissions information is available at www.nature.com/reprints.
The authors declare no competing financial interests. Correspondence and requests for materials should be addressed to
L.M.D. (lmduan@umich.edu).

\newpage

\section{Methods}

\subsection{Experimental setup}

We use a home-built confocal microscopy, with an oil-immersed objective lens
($N.A.=1.49$), to address and detect single NV center in a type IIa
single-crystal synthetic diamond sample (Element Six). A $532$ nm diode
laser, controlled by an acoustic optical modulator (AOM), is used for spin
state initialization and detection. We collect fluorescence photons
(wavelength ranging from $637-850$ nm) into a single-mode fiber and detect
them by the single-photon counting modular (SPCM), with a counting rate $105$
kHz and a signal-to-noise ratio $15:1$. The diamond sample is mounted on a $3
$-axis closed-loop Piezo for sub-micrometer resolution scanning. An
impedance-matched gold coplanar waveguide (CPW) with $70$ $\mu $m gap,
deposited on a cover-glass, is used for delivery of radio-frequency (RF) and
microwave (MW) signals to the NV\ center.

In our experiment, we find a single NV center with a proximal $C^{13}$ of $%
13.7$ MHz hyperfine strength (Fig. 1). To polarize the nearby nuclear spins (%
$C^{13}$ and the host $N^{14}$), we apply a magnetic field of $451$ G along
the NV axis using a permanent magnet. Under this field, the electron spin
levels $\left\vert m=0\right\rangle $ and $\left\vert m=-1\right\rangle $
become almost degenerate in the optically excited state (called the esLAC,
the electron spin level anti-crossing \cite{22}), which facilitates
electron-spin nuclear-spin flip-flop process during optical pumping. The
spin flip-flop process leads to polarization of the nitrogen nuclear spin on
the NV\ site and the nearby $C^{13}$ nuclear spins after $2\mu s$ green
laser illumination \cite{22}. The Zeeman energy from the $451$ G magnetic
field shifts the energy difference between electron spin states $\left\vert
m=0\right\rangle $ and $\left\vert -1\right\rangle $ ($\left\vert
+1\right\rangle $) from the zero-field splitting $2870$ MHz to $1601$ MHz ($%
4141$ MHz) and the nuclear spin hyperfine splitting from $13.7$ MHz to $14.15
$ MHz ($13.25$ MHz) for $\left\vert -1\right\rangle $ ($\left\vert
+1\right\rangle $) levels. Due to the large splitting of $m=\pm 1$ levels,
we apply two independent MW sources (Rohde-Schwarz), locked by a $10$ MHz
reference Rubidium clock, to address each transition. To adjust the
frequency and phase of the MW pulses, we mix each MW output with an AWG
(Tektronix, $500$ MHz sample rate). RF signals for nuclear spin manipulation
are generated directly by another analog channel of the AWG. All the MW and
RF signals are amplified by independent amplifiers, combined through a
home-made circuit, and delivered to the CPW. The digital markers of the AWG
are used to control the pulse sequence (including laser and SPCM) with a
timing resolution of $2$ ns.

For each experimental cycle, we start the sequence with $2$ $\mu s$ laser
illumination to polarize NV electron spin and nearby nuclear spins and end
it with a $3$ $\mu s$ laser pulse for spin state detection. We collect
signal photons for $300$ ns right after the detection laser rises, and
another $300$ ns for reference $2$ $\mu s$ later. With a photon collection
rate of $105$ kHz, we have an average of $0.03$ photon counts per cycle. For
measurement of each data, we repeat the experimental cycle at least $10^{6}$
times, resulting in a total photon counts of $3\times 10^{4}$. The error
bars of our data account for the statistical error associated with the
photon counting. To calculate the error bar of each data, we use Monte Carlo
simulation by assuming a Poissonian distribution for the photon counts. For
each simulation trial, we calculate the value of each data. Then, by sampling
over all the trails according to the Poissonian distribution, we get
statistics of the data (including its mean value and standard deviation,
the error bar).

\subsection{Calibration of fluorescence levels for different states}

Due to the esLAC that induces spin flip-flop during the detection and the
imperfect initial polarization of the electron and nuclear spins, each spin
component $|m,m_{n}\rangle $ ($m=0,\pm 1;$ $m_{n}=\uparrow ,\downarrow $)
may fluorescent at different levels. Note that the spins are dominantly in
the state $|m=0,m_{n}=\uparrow \rangle $ after the optical pumping. To
calibrate the fluorescence level of each state, we therefore associate the
detected fluorescence level right after the optical pumping with the state $%
|m=0,m_{n}=\uparrow \rangle $. With MW or RF $\pi $-pulses (the $\pi $%
-pulses are calibrated through Rabi oscillations), we can make a complete
transfer between $|m=0,m_{n}=\uparrow \rangle $ and any other $%
|m,m_{n}\rangle $ spin component. For instance, with a $\pi $-pulse between $%
|m=0,m_{n}=\uparrow \rangle $ and $|m=0,m_{n}=\downarrow \rangle $ right
after the optical pumping, we associate the detected fluorescence level with
the $|m=0,m_{n}=\downarrow \rangle $ state. In this way, the characteristic
fluorescence level of each component $|m,m_{n}\rangle $ can be calibrated.
With the calibrated fluorescence level for each spin component, we then read
out the system state after the geometric gates through quantum state
tomography \cite{23b}.

\subsection{Quantum Process tomography}

A quantum process can be described by a completely positive map $\varepsilon
$ acting on an arbitrary initial state $\rho _{i}$, transferring it to $\rho
_{f}\equiv \varepsilon (\rho _{i})$. In quantum process tomography (QPT), we
choose a fixed set of basis operators $\{E_{m}\}$ so that the map $%
\varepsilon (\rho _{i})=\sum_{mn}E_{m}\rho _{i}E_{n}^{\dagger }\chi _{mn}$
is identified with a process matrix $\chi _{mn}$.\ We experimentally measure
this process matrix $\chi $ through the maximum likelihood technique \cite%
{23b}. For single-bit QPT, we set the basis operators as $I=I$, $X=\sigma
_{x}$, $Y=-i\sigma _{y}$, $Z=\sigma _{z}$ and choose four different initial
states $|0\rangle $, $|1\rangle $, $(|0\rangle +|1\rangle )/\sqrt{2}$, and $%
(|0\rangle -i|1\rangle )/\sqrt{2}$. We reconstruct the corresponding final
density operators through the standard quantum state tomography and use them
to calculate the process matrix $\chi _{e}$. This process matrix $\chi _{e}$
is compared with the ideal one $\chi _{id}$ by calculating the process
fidelity $F_{P}=Tr(\chi _{e}\chi _{id})$. The process fidelity $F_{P}$ also
determines the average gate fidelity $\overline{F}$ by the formula $%
\overline{F}=(dF_{P}+1)/(d+1)$ \cite{23b}, where $\overline{F}$ is defined
as the fidelity averaged over all possible input states with equal weight
and $d$ is the dimension of the state space (with $d=2$ for a single qubit).

\end{document}